\documentstyle[12pt]{article}

\begin{document}

\textheight 9.0in
\topmargin -0.5in
\textwidth 6.5in
\oddsidemargin -0.01in
\def\singlespace {\smallskipamount=3.75pt plus1pt minus1pt
                  \medskipamount=7.5pt plus2pt minus2pt
                  \bigskipamount=15pt plus4pt minus4pt
                  \normalbaselineskip=15pt plus0pt minus0pt
                  \normallineskip=1pt
                  \normallineskiplimit=0pt
                  \jot=3.75pt
                  {\def\smallskip {\vskip\smallskipamount}}
                  {\def\medskip   {\vskip\medskipamount}}
                  {\def\bigskip   {\vskip\bigskipamount}}
                  {\setbox\strutbox=\hbox{\vrule 
                    height10.5pt depth4.5pt width 0pt}}
                  \parskip 7.5pt
                  \normalbaselines}
\def\middlespace {\smallskipamount=5.625pt plus1.5pt minus1.5pt
                  \medskipamount=11.25pt plus3pt minus3pt
                  \bigskipamount=22.5pt plus6pt minus6pt
                  \normalbaselineskip=22.5pt plus0pt minus0pt
                  \normallineskip=1pt
                  \normallineskiplimit=0pt
                  \jot=5.625pt
                  {\def\smallskip {\vskip\smallskipamount}}
                  {\def\medskip   {\vskip\medskipamount}}
                  {\def\bigskip   {\vskip\bigskipamount}}
                  {\setbox\strutbox=\hbox{\vrule 
                    height15.75pt depth6.75pt width 0pt}}
                  \parskip 11.25pt
                  \normalbaselines}
\def\doublespace {\smallskipamount=7.5pt plus2pt minus2pt
                  \medskipamount=15pt plus4pt minus4pt
                  \bigskipamount=30pt plus8pt minus8pt
                  \normalbaselineskip=30pt plus0pt minus0pt
                  \normallineskip=2pt
                  \normallineskiplimit=0pt
                  \jot=7.5pt
                  {\def\smallskip {\vskip\smallskipamount}}
                  {\def\medskip   {\vskip\medskipamount}}
                  {\def\bigskip   {\vskip\bigskipamount}}
                  {\setbox\strutbox=\hbox{\vrule 
                    height21.0pt depth9.0pt width 0pt}}
                  \parskip 15.0pt
                  \normalbaselines}

\begin{center}
{\bf {\Large Divergence of the Quantum Stress Tensor on}}

{\bf {\Large the Cauchy Horizon in 2-d Dust Collapse}}

\bigskip\ 

{\bf Sukratu Barve \footnote{{e-mail address: sukkoo@relativity.tifr.res.in}}
 and T. P. Singh\footnote{{ e-mail address:
tpsingh@tifr.res.in}}}

{\it Tata Institute of Fundamental Research,}

{\it Homi Bhabha Road, Mumbai 400 005, India.}

\smallskip\ 

{\bf Cenalo Vaz\footnote{{ e-mail address: cvaz@ualg.pt}}}

{\it Unidade de Ciencias Exactas e Humanas}

{\it Universidade do Algarve, Faro, Portugal}

\bigskip

\end{center}

\begin{abstract}
We prove that the quantum stress tensor for a massless scalar field in two
dimensional non-self-similar Tolman-Bondi dust collapse and Vaidya radiation
collapse models diverges on the Cauchy horizon, if the latter exists. The two
dimensional model is obtained by suppressing angular coordinates in the
corresponding four dimensional spherical model.
 
\end{abstract}

\middlespace

\section{Introduction}

If a classical model of gravitational collapse results in the formation of a
naked singularity, quantum effects can be expected to play a significant
role during the final stages of the collapse. One way to study these effects
is the quantization of test matter fields in the background spacetime
provided by the collapsing classical matter. The semiclassical 
approximation is expected to be valid up to Planck scales.
In particular, one is
interested in the behavior of the stress tensor of the quantized field in
the approach to the Cauchy horizon. The divergence of the vacuum expectation
value of the quantized stress tensor on the Cauchy horizon signals an
instability of the horizon, and suggests that back-reaction will prevent the
naked singularity from forming.

Two well-known examples of formation of naked singularities are the
spherical collapse of inhomogeneous dust (the Tolman-Bondi model) and the
spherical collapse of null dust (the Vaidya model). It is known for both
these models that for certain initial data the collapse ends in a black hole
and for other initial data it ends in a naked singularity \cite{tol}, \cite
{his}, \cite{vai}. The stress-tensor of a quantized scalar field on these
background spacetimes has been investigated, by specializing to the case of
2-d self-similar collapse. The restriction to 2-d is 
similar to the geometric optics approximation in 4-d - the latter amounts 
to keeping only the $l=0$ modes.
The 2-d spacetime is obtained by suppressing the
angular coordinates in the 4-d spherical model. For a two dimensional model,
explicit expressions for the vacuum expectation value of the stress-tensor
can be obtained from the trace anomaly, by imposing conservation of the
stress tensor. The assumption of self-similarity allows double null
coordinates to be constructed explicitly. Using these coordinates it has
been shown that the outgoing quantum flux diverges on the Cauchy horizon,
for self-similar Vaidya collapse \cite{his}, as well as for self-similar
Tolman-Bondi collapse \cite{bar}.

A priori, it may be the case that the divergence in the 2-d model could be
because of the assumption of self-similarity. In this paper, we prove that
this assumption can be relaxed, and that the outgoing quantum flux will
diverge on the Cauchy horizon, for all initial conditions for which a naked
singularity forms in the 2-d Vaidya and Tolman-Bondi models.

The outline of the proof is as follows. Consider the collapse of a classical
non-self-similar spherical dust cloud and choose the initial conditions to
be such that the collapse results in a naked singularity. We now construct a
new initial distribution by replacing a spherical region by a self-similar
distribution. The new distribution is hence a self-similar spherical region
surrounded by part of the original distribution. The free parameter of the
self-similar distribution is fixed by requiring the first and second
fundamental forms to match at the boundary between the self-similar region
and the original distribution. It is then shown that if the evolution of the
original distribution results in a naked (covered) singularity, the
evolution of the modified distribution also results in a naked (covered)
singularity. We show that, in general, the density of the cloud will change
discontinuously at the boundary, but this change will be finite, and not
infinite.

We next consider the quantum stress tensor for a massless scalar field on
the classical background dust spacetime. As has been shown earlier, the
stress tensor diverges on the Cauchy horizon for a self-similar model.
Consider now the modified distribution (self-similar region surrounded by
non-self-similar region) mentioned above. We show that the nature of the
divergence in the self-similar region is such that it implies a divergence
in the outer non-self-similar region as well. Then, by considering a family
of initial distributions, the size of the self-similar region is shrunk to
zero - for each distribution in the family there is a divergence on the
Cauchy horizon. The limiting distribution, in which the self-similar region
disappears entirely, is the original distribution, and this also has a
divergence on the Cauchy horizon.

The plan of the paper is as follows. In Section 2 we construct the modified
distribution which includes a self-similar spherical region in the interior.
In Section 3 we obtain the quantum stress tensor for a test scalar field
on this modified
distribution and prove that it diverges on the Cauchy horizon for a
non-self-similar dust cloud.

\section{The Modified Distribution}

In this Section, we show how the modified initial distribution is
constructed, for the cases of marginally bound and non-marginally bound dust
collapse, as well as for the Vaidya model.

\subsection{Marginally Bound Dust Collapse}

The collapse of a spherical dust cloud is described by the Tolman-Bondi
line-element, using comoving coordinates $(t,r,\theta ,\phi )$, 
\begin{equation}
\label{ele}ds^2=-dt^2+\frac{R^{\prime 2}}{1+f(r)}dr^2+R^2(t)d\Omega ^2, 
\end{equation}
where $R(t,r)$ is the area radius at time $t$ of the shell labeled $r$, and 
$f(r)$ is a free function, satisfying $f>-1$. The marginally bound solution
is one for which $f(r)=0$. The only non-zero component of the
energy-momentum tensor is the energy density $\rho (t,r),$ which satisfies
the Einstein equation 
\begin{equation}
\label{rho}\rho =\frac{F^{\prime }(r)}{R^2R^{\prime }} 
\end{equation}
where $F(r)$ is another free function, and has the interpretation of being
the mass to the interior of the shell $r$. The only other Einstein equation
is 
\begin{equation}
\label{ei2}\dot R^2=\frac{F(r)}R+f(r). 
\end{equation}

Let us consider the collapse of marginally bound dust cloud, starting at a
time $t_i$ and having an initial density distribution $\rho _0(r)$, for $%
r\leq r_b$, where $r_b$ is the boundary of the star. Integrating (\ref{ei2})
gives the solution 
\begin{equation}
\label{sol}R^{3/2}(t,r)=\frac 32\sqrt{F(r)}\left( t_0(r)-t\right) 
\end{equation}
for the evolution of the area radius of the shell $r$. $t_0(r)$ is a
function of integration, to be determined by choosing an initial scaling, $%
R(t_i,r)$, at the start of collapse. The area shrinks to zero at $t=t_0(r)$,
resulting in the formation of a curvature singularity. It is the singularity
at $r=0$, the central singularity, which is of interest to us, as this has
been shown to be naked for some initial conditions. Specifically, it has
been shown \cite{tol} that if the initial density distribution $\rho _0(R)$
has a Taylor expansion 
\begin{equation}
\label{exp}\rho _0(R)=\rho _0+\rho _1R+
{1\over 2}\rho _2R^2+{1\over 6}\rho _3R^3+... 
\end{equation}
then the singularity is at least locally naked if one of the following
conditions is satisfied: (i) $\rho _1<0$, or (ii) $\rho _1=0,\rho _2<0$, or
(iii) $\rho _1=\rho _2=0,\rho _3<0$ and 
$\xi =\sqrt{3}\rho _3/4\rho _0^{5/2}$ is less
than or equal to $-25.9904.$ The singularity may or may not be globally
naked. We are interested in showing that in either case the outgoing quantum
flux diverges on the Cauchy horizon.

The self-similar solution, i.e. one for which the spacetime of the
collapsing cloud possesses a homothetic Killing vector field, is a special
case of the marginally bound solution. If we choose the scaling in such a
way that $t_0(r)=r$, then the mass function of the self-similar solution is
of the form $F_{ss}(r)=\lambda r$, with $\lambda $ a non-negative constant.
All dimensionless quantities are functions of $t/r$. The central singularity
forms at $t=0$, and is known to be (globally) naked for $\lambda \leq
\lambda _c = 0.1809$ \cite{op}. It can also be shown that the initial density
distribution for a self-similar cloud is of the form 
\begin{equation}
\label{sel}\rho _0(R)=\rho _0+\rho _3R^3+\rho _6R^6+\rho _9R^9+... 
\end{equation}

In order to construct the modified initial density distribution, we start
with the original distribution $\rho _0(r)$ and replace a central region, up
to some $r=r_c$, by the self-similar solution (with $r_c<r_b$). For $%
r_c<r<r_b$, the distribution continues to be the original distribution $\rho
_0(r)$. There is only one free parameter in the self-similar solution,
namely $\lambda $, and this is determined by requiring that the total mass
in the self-similar region is equal to the total mass contained in the
original distribution, up to $r=r_c$. That is, $\lambda r_c=F(r_c)$.

We would now like to ensure that the determined value of $\lambda $ is such
that in the limit $r_c$ going to zero, the modified distribution admits a
naked singularity if and only if the original distribution $\rho _0(r)$
admits a naked singularity. This is essential because we are interested in
examining properties of the Cauchy horizon in the original distribution.
Hence it is natural to demand that the modified distribution possess a
Cauchy horizon if and only if the original distribution does. The limiting
value $\lambda _0$, of $\lambda ,$ is clearly $dF(r)/dr|_{r=0}$, and should
satisfy the condition $\lambda \leq \lambda _c$ if and only if the original
distribution $\rho _0(r)$ admits a naked central singularity.

To check this, we go back to the solution (\ref{sol}) and assume, for
simplicity, the scaling to be such that $t_0(r)=r$, so that the solution can
be written as 
\begin{equation}
\label{rso}R^3=\frac 94F(r)\,\left( t-r\right) ^2. 
\end{equation}
We substitute this in (\ref{rho}) and assuming the collapse to begin at $%
t=t_{in}$, find the initial density, $\rho _0(r)$, near the center, to be 
\begin{equation}
\label{rin}\rho _0(r)=\frac 4{3t_{in}^2}\left[ 1+\frac{2F(r)}{F^{\prime
}(r)\,t_{in}}\right] ._{} 
\end{equation}
We compare this with the form (\ref{exp}) for the initial density, after
using (\ref{rso}) with $t=t_{in}$ in (\ref{exp}). This comparison allows us
to deduce the form of $F(r)$ near $r=0$, from which the limiting value $%
\lambda _0=dF(r)/dr|_{r=0}$ can be worked out. We find that for two of the
naked cases, (i) $\rho _1<0$, and (ii) $\rho _1=0,\rho _2<0$, we get $%
\lambda _0=0$, which is a special case of the naked self-similar model. For
the case (iii) $\rho _1=\rho _2=0,\rho _3<0$ we get 
$\lambda _0^{3/2}=-8\rho_{0}^{5/2}/\sqrt{3}\rho_{3}$, which
implies that the chosen self-similar distribution is naked if and only if
the original distribution of type (iii) is naked. For the covered case,
which is $\rho _1=\rho _2=\rho _3=0$, we get that $\lambda _0$ is infinite,
which is a covered self-similar distribution. Hence it is shown that the
modified distribution admits a naked singularity if and only if the original
distribution does.

Next, we discuss the question of the matching of the self-similar region and
the non-self-similar region, at the boundary $r_c$. By comparing the metrics
of the two spacetimes at this boundary we find that the area radii in the
self-similar metric and in the non-self-similar one will be equal if the two
mass functions are equal. Since the masses have been chosen to be equal,
that ensures the matching of the first fundamental form (i.e. the line
element on the boundary). The second fundamental form (i.e. the extrinsic
curvature) is defined to be 
\begin{equation}
\label{ext}K_{\mu \nu }=-\frac 12\left( \xi _{\alpha ;\beta }+\zeta _{\beta
;\alpha }\right) h_\mu ^\alpha h_\nu ^\beta 
\end{equation}
where $\xi _\alpha ^{}$ is a unit normal to the boundary, and $h_\beta
^\alpha =\delta _\beta ^\alpha $ $-\xi ^\alpha \xi _\beta $ is the
projection tensor. The only non-zero component of the extrinsic curvature is 
$K_\theta ^\theta =K_\phi ^\phi =-1/R$, which matches for the two metrics,
since the mass functions have been chosen to be equal. Hence we are assured
that the first and second fundamental forms match on the boundary.

We note that the derivative $F^{\prime }(r)$ will in general not match at
the boundary, when calculated for the self-similar region, and for the
non-self-similar region. As a result, the density function $\rho (t,r)$,
given by (\ref{rho}), will in general be discontinuous at the boundary.
However, it is important for our purposes to note that this discontinuity
will be finite, since $F^{\prime }(r)$ will be finite.

\subsection{Non-marginally Bound Dust Collapse}

The metric function $f(r)$ which appears in (\ref{ele}) is now non-zero.
Equation (\ref{ei2}) can be solved exactly, for given $F(r)$ and $f(r)$.
Whereas $F(r)$ is determined from the initial density distribution, $f(r)$
gets determined from the velocity distribution $\dot R$ at the onset of
collapse.

Given an initial distribution of density and velocity which is non-marginal,
one cannot straightaway match it to a self-similar region at a boundary $r_c$%
, because for the former distribution we have $f(r)\neq 0$, and for the
latter we have $f(r)=0$. Matching of the line-elements requires that both $%
F(r)$ and $f(r)$ match on the boundary between the two regions.

Thus, given the original non-marginal distribution ($\rho _0(r),f(r)$) we
construct the modified distribution as follows. Replace the original model
by a self-similar one, as before, for $r\leq r_c$. For some $r_{*}<r_b$,
introduce a non-marginal dust distribution $(\rho _0(r),f_s(r)$) in the
region $r_c<r<r_{*}$, which has the property that $\rho _0(r)$ (and hence $%
F(r)$) is the same as in the original distribution; $f_s(r_c)=0$ and $%
f_s(r_{*})=f(r_{*})$. As before, the self-similar model is chosen such that $%
\lambda r_c=F(r_c)$. The introduction of this sandwiched region ensures that
the first and second fundamental forms are matched at the boundaries $r_c$
and $r_{*}$. The limiting value of $\lambda $, as $r_c$ goes to zero, is
again given by $\lambda _0=dF(r)/dr|_{r=0}$, and is calculated by letting
both $r_c$ and $r_{*}$ go to zero, while always keeping $r_{*}>r_c$. By
carrying out an analysis similar to the marginal case, it can be shown that $%
\lambda _0$ is such that the introduced self-similar model is naked if and
only if the original distribution is naked.

\subsection{The Vaidya Model}

The collapse of a null dust cloud is described by the Vaidya metric 
\begin{equation}
\label{vai}ds^2=-\left( 1-\frac{2m(v)}R\right) dv^2+2dvdR+r^2d\Omega ^2 
\end{equation}
where the mass function $m(v)$ depends on the advanced time coordinate $v$.
The mass function is zero for $v<0$, and constant for $v>v_b$. Thus the
cloud is bounded in the region $0<v<v_b$, and its exterior is Schwarzschild
spacetime. A curvature singularity forms when the inner boundary of the
cloud hits $R=0$. This singularity is known to be naked for $%
dm(v)/dv|_{v=0}\leq 1/16$ \cite{vai} and covered for $dm(v)/dv|_{v=0}>1/16$.
The special case of a self-similar model is described by the linear mass
function, $m(v)=\mu v$, $0<v<v_b$.

Given a non-self-similar model $m(v)$, we construct the modified
distribution by replacing the original model by a self-similar one in the
region $0<v<v_c$, with $v_c<v_b$. The free parameter $\mu $ in the
self-similar model is fixed by demanding $\mu v_c=m(v_c).$ The limiting
value of $\mu $, as $v_c$ goes to zero, is $dm(v)/dv|_{v=0}$, and it is
apparent that the introduced self-similar distribution is naked if and only
if the original distribution $m(v)$ is naked.

Since the boundary $v_c$ is a null hypersurface, the matching of the
spacetimes at the boundary is done by comparing, not the line elements, but
the affine parameter for outgoing or ingoing null geodesics.
We now show that the matching of the affine parameters
on the boundary $v_c$ implies that the two mass functions should be equal.
For this purpose, it is convenient to restrict to the 2-d line element
obtained from (\ref{vai}) by suppressing the angular coordinates, and to
write it in double null coordinates $u,v$, i.e. 
\begin{equation}
\label{dnu}ds^2=C^{2}(u,v)\,du\,dv. 
\end{equation}
The function $C(u,v)$ satisfies the differential equation 
\begin{equation}
\label{see}\frac{C^2,v}{C^4}\left( 1-\frac{2m(v)}R\right) +\frac{2m(v)}{%
R^2C^2}+2\frac{C^2,v}{C^4}=0. 
\end{equation}
By integrating the geodesic equation for the metric (\ref{dnu}) it is shown
that the affine parameter along ingoing and outgoing null geodesics is of
the form 
\begin{equation}
\label{aff}p=a\int C^2du+b\quad ,\quad q=c\int C^2dv+d. 
\end{equation}
Matching of the affine parameter at the boundary $v_c$ hence requires that $%
C_a^2du=C_b^2du$, and the use of this equality in the differential equation (%
\ref{see}) gives the result that the mass functions of the two regions
(self-similar and non-self-similar) must match. Further, it may be shown, as
in the dust case, that the extrinsic curvatures match at the boundary
between the two regions.

We have now completed the construction of the modified classical
distribution, which will be used to prove the divergence of the outgoing
quantum flux on the Cauchy horizon.

\section{\bf The Quantum Stress Tensor}

We restrict attention to the two dimensional spacetime (Tolman-Bondi or
Vaidya) obtained by suppressing angular coordinates in the four-dimensional
spherical spacetime. The expectation value $<0_{in}|T_{\mu \nu }|0_{in}>$ of
the energy-momentum tensor of a quantized scalar field in the Minkowski
vacuum $|0_{in}>$ can be calculated from the trace anomaly, and Wald's
axioms. The trace anomaly is equal to ${\cal R}/24\pi $, where ${\cal R}$ is
the Ricci scalar for the background spacetime \cite{dav}. The two
dimensional spacetime can be expressed in terms of global null coordinates $%
\hat u$ and $\hat v$ as 
\begin{equation}
\label{dou}ds^2=C^2(\hat u,\hat v)\,d\hat u\,d\hat v. 
\end{equation}
These coordinates are chosen such that the initial quantum state of the
scalar field , which is the standard Minkowski vacuum $|0_{in}>$ on 
${\cal I}^{-}$, is the vacuum with respect to the normal modes of the scalar 
wave
equation in $\hat u,\hat v$ coordinates. The components of $<T_{\mu \nu }>$
are given by

\begin{equation}
\label{mumu}<T_{\hat u\hat u}>=-\frac 1{12\pi }C\left( \frac 1C\right)
_{,\hat u,\hat u}{,} 
\end{equation}
\begin{equation}
\label{vv}<T_{\hat v\hat v}>=-\frac 1{12\pi }C\left( \frac 1C\right) _{,\hat
v,\hat v}{,} 
\end{equation}
\begin{equation}
\label{muvi}<T_{\hat u\hat v}>=\frac{{\cal R}C^2}{96\pi }. 
\end{equation}

Consider a situation where two solutions for the background metric 
 are matched across a hypersurface, like a geodesic, for instance,
 which in fact will be the case for our models. Let $C_{1}^{2}$ and
 $C_{2}^{2}$ be the metrics in the two regions. In either of the regions
 where the solutions are given, the expressions 
(\ref{mumu}-\ref{muvi}) are 
analytic.
Now consider the hypersurface (boundary) where the matching has taken
 place. Assuming the line element on the boundary and the second fundamental 
form of
the boundary being calculated in both of the regions to be identical
 respectively, one obtains
 $
 C_{1}=C_{2}=C
 $
at the boundary.
 The difference in the quantum stress tensor components can be 
 expressed at the boundary ( e.g. from equation 1).

 \begin{equation}
\label{dif}
 <T_{1\hat{u}\hat{u}}>-<T_{2\hat{u}\hat{u}}>=
 -1/12\pi C\left(1/C_{1}-1/C_{2}\right)_{,\hat{u},\hat{u}}
 \end{equation}
 Since both the metrics are analytic in their respective regions (we only
 require their second order partial derivatives to be finite) and if
 one extends each one of them smoothly across the boundary, one finds that
 the expression above is certainly finite.
 So, the discontinuous change in $<T_{\hat{u}\hat{u}}>$ across the boundary
 will be finite. The same can be said of the rest of the components.

We next introduce double null coordinates in the various regions of the
modified distribution under consideration. Consider first the case of
marginally bound dust. The various regions are the introduced self-similar
region (henceforth labeled $2$) in the coordinate range $0<r<r_c$, the
original distribution in the region (henceforth labeled $1$) 
$r_{c}<r<r_b$, and
the Schwarzschild region (henceforth labeled $0)$ $r>r_b$. We write the
line-elements in these regions as 
\begin{equation}
\label{too}ds^2=A^2(u_2,v_2)du_2dv_2 
\end{equation}
in region $2$, as 
\begin{equation}
\label{won}ds^2=B^2(u_1,v_1)du_1dv_1 
\end{equation}
in region $1,$ and finally, as 
\begin{equation}
\label{zer}ds^2=D^2(u,v)dudv 
\end{equation}
in region $0$.

The relationship amongst these coordinates can be given, following the
procedure described in Birrel and Davies \cite{bir}. Because of matching the
first fundamental form at various boundaries, we have 
\begin{equation}
\label{uoo}u_1=\alpha _1(u),\quad v=\beta _1(v_1), 
\end{equation}
and 
\begin{equation}
\label{utw}u_2=\alpha _2(u_1),\quad v_1=\beta _2(v_2). 
\end{equation}

The center of the cloud being given by both 
\begin{equation}
\label{cen}\hat u=\hat v 
\end{equation}
and 
\begin{equation}
\label{ce2}u_2=v_2-2R_0 
\end{equation}
in a manner similar to Birrell and Davies, we obtain a relation between $%
\hat u$ and $u_1$ 
\begin{equation}
\label{re1}\hat u=\beta _1\left( \beta _2\left( \alpha _2\left( u_1\right)
+2R_0\right) \right) 
\end{equation}
as well as a relation between $\hat u$ and $u$ 
\begin{equation}
\label{re2}\hat u=\beta _1\left( \beta _2\left( \alpha _2\left( \alpha
_1\left( u\right) \right) +2R_0\right) \right) . 
\end{equation}

Starting from the expressions 
(\ref{mumu}-\ref{muvi}) for the expectation value of the
stress-tensor, and using the coordinate relationships given above, we obtain
the following expressions for the $T_{\mu\nu}$ in various regions. In region
2, 
\begin{equation}
\label{t11}<T_{u_2u_2}>=-F_{u_2}\left( A^2\right) +F_{u_2}\left( \frac{d\hat
u}{du_2}\right) ,
\end{equation}
\begin{equation}
\label{t12}<T_{v_2v_2}>=-F_{v_2}\left( A^2\right) +F_{v_2}\left( \frac{d\hat
v}{dv_2}\right) ,
\end{equation}
\begin{equation}
\label{t13}<T_{u_2v_2}>={1\over 6\pi A^2}
\left( \frac{d\hat u}{du_2}\frac{d\hat v}{dv_2}%
\right) ^2\left[ \frac{A_{,u_2,v_2}^2}{A^2}-\frac{%
A_{,u_2}^2A_{,v_2}^2}{A^4}\right]. 
\end{equation}

The function $F_{x}(y)$ is defined as

\begin{equation}
\label{fex} F_{x}(y)={1\over 12\pi} \sqrt{y} \left( 1\over 
\sqrt{y}\right)_{,x,x}
\end{equation}

In region 1,

\begin{equation}
\label{t1a}<T_{u_1u_1}>=-F_{u_1}\left( B^2\right) +F_{u_1}\left( \frac{d\hat
u}{du_1}\right) ,
\end{equation}
\begin{equation}
\label{t1b}<T_{v_1v_1}>=-F_{v_1}\left( B^2\right) +F_{v_1}\left( \frac{d\hat
v}{dv_1}\right) ,
\end{equation}
\begin{equation}
\label{t1c}<T_{u_1v_1}>= {1\over 6\pi B^2}
\left( \frac{d\hat u}{du_1}\frac{d\hat v}{dv_1}%
\right) ^2\left[ \frac{B_{,u_1,v_1}^2}{B^2}-\frac{%
B_{,u_1}^2B_{,v_1}^2}{B^4}\right]. 
\end{equation}

Similarly in region 0, 
\begin{equation}
\label{t01}<T_{uu}>=F_u\left( D^2\right) +\alpha ^{\prime }F_{u_2}\left(
\beta ^{\prime }\right) +F_u\left( \alpha ^{\prime }\right), 
\end{equation}
\begin{equation}
\label{t02}<T_{vv}>=-F_v\left( D^2\right) ,
\end{equation}
\begin{equation}
\label{t03}<T_{uv}>=-1/24\pi \left( ln\left( D^2\right) \right) _{,u,v}, 
\end{equation}

where $^{\prime }$ indicates first derivative with respect to the argument,
and

\begin{equation}
\label{alb}\alpha \left( {}\right) =\alpha _2\left( \alpha _1\left(
{}\right) \right) ,\qquad \beta \left( {}\right) =\beta _1\left( \beta
_2\left( {}\right) \right) . 
\end{equation}

Let us first consider region 2; we have the self-similar metric there
according to the procedure adopted. Consider a parameter value $\lambda $
for which the collapse results in a naked central singularity. From Barve et
al. \cite{bar} we know that the component $<T_{u_2u_2}>$ diverges on the
entire Cauchy horizon in the region (the other components remaining finite).
It is obvious that $<T_{\hat u\hat u}>$ will also diverge there. As
pointed out above, it is
important now to note that the discontinuity in the quantum stress tensor,
at the boundary $r_c,$ is finite at the boundary between regions 2 and 1.
This is because the various expectation values are determined by the trace
anomaly and derivatives (up to second derivative) of the metric. The metric
derivatives are finite on the Cauchy horizon, for $r>0$. Hence the quantum
stress tensor component $<T_{u_1u_1}>$ at the boundary, in the limit of
approach from region 1, will be divergent.

We now examine the expression for $<T_{u_1u_1}>$ , i.e. equation (\ref{t1a}%
). The first term in this expression is finite, because we demand the metric
component to be at least ${C}^2$. This term is finite not only at the event
on the boundary but all over the null ray (Cauchy horizon) emanating from
that event (in fact, all over the region 1).

Now, the second statement above implies that the second term in equation (%
\ref{t1a}) diverges at the intersection of the boundary with the null ray
(Cauchy horizon). This term is a function of only the retarded null
coordinate $u_1$. Since it diverges at one event (on the boundary), it
diverges all along the outgoing null ray ($u_1=$ constant). The first
term, as mentioned before, is finite there. Hence the tensor component $%
<T_{u_1u_1}>$ diverges all along the Cauchy horizon in region 1 as well.

Finally, we consider a family of modified distributions, each with a
successively smaller value of the boundary coordinate $r_c$, so that in the
limiting case, $r_c$ tends to zero. For each family in the distribution,
there is a divergence of the outgoing flux, in region 1. In the limit that $%
r_c$ tends to zero, we recover the original non-self-similar distribution $%
\rho _0(r)$ which admits a naked singularity, and by virtue of the
construction given here, has a divergence of $<T_{u_1u_1}>$ on the Cauchy
horizon.

We can make a similar argument at the second boundary, between region 1 and
the Schwarzschild region 0. Thus, the divergence occurs all over the Cauchy
horizon, to whichever extent it exists. In particular, if we have a globally
naked singularity, there is a divergence at the intersection of the Cauchy
horizon with ${\cal I}^{+}$.

Also, we note that this argument works for any number of regions with
different metric solutions matched at the boundaries. This is important from
the point of view of the extra shell we needed to introduce in the
non-marginally bound Tolman Bondi case. The finiteness of the rest of the
tensor components can be argued on similar lines.

As regards the Vaidya metric, the entire argument is the same except for the
fact that the boundaries between the regions are ingoing null geodesics. In 
fact, $\beta ()$ being the identity function \cite{his}, the calculation is
simplified considerably.

\section{\bf Conclusion}

We have shown that the quantum stress tensor diverges on the Cauchy horizon,
if it exists, in non-self-similar Tolman Bondi dust and Vaidya radiation
collapse in two dimensions. There is no direct way of deducing this by
analytical calculations of the expressions in the general case. However, we
employ a limiting process of approaching the required spacetime metric via
patching up tractable solutions. From this technique, it appears that the
divergence technically results from the metric rendered non-invertible at
the Cauchy horizon in the self-similar portion. This does not seem to be the
ultimate reason for the divergence, for we see that the divergence persists
even after the limit to the actual metric is taken. Needless to say, the
divergence does not seem to be ultimately a result of the self-similar
nature although that does help in making the divergence evident in the
calculations.

\end{document}